\documentclass[conference]{IEEEtran}
\ifCLASSINFOpdf
\else
\fi
\usepackage{amsmath,dsfont,bbm,epsfig,amssymb,amsfonts,amstext,verbatim}\usepackage{amsopn,cite,subfigure}
\usepackage{multirow,multicol,lipsum,xfrac}
\usepackage{amsthm}
\usepackage{mathtools,amsthm}
\usepackage{perpage}
\usepackage{balance}
\usepackage{url}
\usepackage{amsfonts}
\usepackage{epsfig}
\usepackage[font={small}]{caption}
\usepackage{psfrag}	
\usepackage{etoolbox}
\usepackage{algorithmicx}
\usepackage[Algorithm,ruled]{algorithm}
\usepackage{algpseudocode}
\usepackage{pifont}
\usepackage[utf8]{inputenc}
\usepackage[T1]{fontenc}  
\usepackage[nolist]{acronym}
\MakePerPage{footnote}
\usepackage{paralist}
\usepackage{enumitem}
\usepackage{bbm}
\usepackage[process=auto]{pstool}
\usepackage{tikz}
\usetikzlibrary{shapes,arrows}

\newcommand{\setL}{\mathbb{L}}

\newcommand{\setS}{\mathbb{S}}

\newcommand{\setC}{\mathbb{C}}

\newcommand{\rms}{\mathrm{s}}

\newcommand{\her}{\mathsf{H}}

\newcommand{\mar}{\mathcal{R}}

\newcommand{\man}{\mathcal{N}}

\newcommand{\mac}{\mathcal{C}}

\newcommand{\mm}{\mathrm{m}}
\newcommand{\ee}{\mathrm{e}}

\newcommand{\bh}{{\mathbf{h}}}

\newcommand{\bx}{{\boldsymbol{x}}}

\newcommand{\set}[1]{\left\lbrace#1\right\rbrace}

\newcommand{\bz}{{\boldsymbol{z}}}

\newcommand{\bg}{{\mathbf{g}}}
\newcommand{\bs}{{\boldsymbol{s}}}

\newcommand{\bw}{{\boldsymbol{w}}}

\newcommand{\by}{{\boldsymbol{y}}}

\newcommand{\bn}{{\boldsymbol{n}}}

\newcommand{\trp}{\mathsf{T}}

\newcommand{\mW}{\mathbf{W}}

\newcommand{\mI}{\mathbf{I}}

\newcommand{\mG}{\mathbf{G}}

\newcommand{\mH}{\mathbf{H}}

\newcommand{\E}{\mathsf{E}\hspace{.5mm}}

\newcommand{\argmax}{\mathop{\mathrm{argmax}}} 

\newcommand{\mh}{\mathbf{h}}

\newcommand{\mw}{\mathbf{w}}

\newcommand{\norm}[1]{\lVert #1 \rVert}
\newcommand{\re}[1]{\mathsf{Re}\left\lbrace #1 \right\rbrace}

\newcommand{\abs}[1]{\lvert #1 \rvert}
\newcommand{\tr}[1]{\mathrm{tr} \{ #1 \}}

\newtheoremstyle{mystyle}
  {}
  {}
  {}
  {}
  {\bfseries}
  {:}
  { }
  {}

\theoremstyle{mystyle}

\algnewcommand\algorithmicLet{\textbf{Let}}
\algnewcommand\Let{\item[\algorithmicLet]}
\algnewcommand\algorithmicSet{\textbf{Set}}
\algnewcommand\Set{\item[\algorithmicSet]}

\algnewcommand\algorithmicInitiate{\textbf{Initiate}}
\algnewcommand\Initiate{\item[\algorithmicInitiate]}
\algnewcommand\algorithmicStart{\textbf{Begin}}
\algnewcommand\Begin{\item[\algorithmicStart]}
\algnewcommand\algorithmicEnd{\textbf{End}}
\algnewcommand\End{\item[\algorithmicEnd]}

\algnewcommand\algorithmicOutP{\textbf{Output:}}
\algnewcommand\Out{\item[\algorithmicOutP]}

\algnewcommand\algorithmicInP{\textbf{Input:}}
\algnewcommand\In{\item[\algorithmicInP]}
\newcommand\NoDo{\renewcommand\algorithmicdo{}}

\newcounter{bar}

\begin{document}
\title{Iterative Antenna Selection for Secrecy Enhancement in Massive MIMO Wiretap Channels}

\author{
\IEEEauthorblockN{
Ali Bereyhi\IEEEauthorrefmark{1}, 
Saba Asaad\IEEEauthorrefmark{1}, 
Rafael F. Schaefer\IEEEauthorrefmark{2} and
Ralf R. M{\"u}ller\IEEEauthorrefmark{1}
}
\IEEEauthorblockA{
\IEEEauthorrefmark{1}Institute for Digital Communications (IDC), Friedrich-Alexander Universit\"at Erlangen-N\"urnberg (FAU)\\
\IEEEauthorrefmark{2}Information Theory and Applications Chair, Technische Universität Berlin (TUB)\\
ali.bereyhi@fau.de, saba.asaad@fau.de, rafael.schaefer@tu-berlin.de, ralf.r.mueller@fau.de
\thanks{This work was supported by the German Research Foundation, Deutsche Forschungsgemeinschaft (DFG), under Grant No. MU 3735/2-1.}
}
}
\IEEEspecialpapernotice{(Invited Paper)}
\IEEEoverridecommandlockouts

\maketitle

\begin{abstract}
The growth of interest in massive MIMO systems is accompanied with hardware cost and computational complexity. Antenna selection is an efficient approach to overcome this cost-plus-complexity issue which also enhances the secrecy performance in wiretap settings. Optimal antenna selection requires exhaustive search which is computationally infeasible for settings with large dimensions. This paper develops an iterative algorithm for antenna selection in massive multiuser MIMO wiretap settings. The algorithm takes a stepwise approach to find a suitable subset of transmit antennas. Numerical investigations depict a significant enhancement in the secrecy performance.\vspace*{3mm} 
\end{abstract}


\IEEEpeerreviewmaketitle

\section{Introduction}
Motivated by emerging increasing traffic demands as well as multi antenna devices and terminals, physical layer security in \ac{mimo} wiretap channels has drawn significant attentions from information-theoretic points of view \cite{khisti2010secure,oggier2011secrecy,liu2009note}. The investigations have demonstrated the promising secrecy performance of these settings and depicted that the growth in system dimensions can significantly boost this performance \cite{kapetanovic2015physical}. Such large-scale setups however suffer from high \ac{rf} cost and~computational~complexity. Therefore, classical approaches such as antenna selection \cite{molisch2005capacity,asaad2017tas}, load modulated arrays \cite{sedaghat2016load} and hybrid analog-digital precoding \cite{liang2014low} have been proposed to alleviate this issue.

The idea of antenna selection is to transmit or receive via a subset of available antennas. By proper selection of the subset, this approach can provide significant advantages in terms of the overall \ac{rf} cost and hardware~complexity~without significant degradation in the performance \cite{sanayei2004antenna,bereyhi2017asymptotics,bereyhi2017nonlinear}. The optimal approach for antenna selection requires an exhaustive search which is computationally impractical particularly in massive \ac{mimo} settings \cite{marzetta2010noncooperative}. Hence, there are several studies devoted to find sub-optimal greedy algorithms with polynomial complexity; see for example \cite{bereyhi2017asymptotics,bereyhi2018stepwise} and references therein for some recent studies. 

Antenna selection is a special case of the general problem of subset selection arising in several applications such as pattern classification \cite{duda2012pattern} and data mining \cite{han2011data}. An efficient low-complexity approach in these applications is stepwise regression in which the selected subset is iteratively constructed such that the growth of a given metric is maximized in each step. Although this strategy does not necessarily result in the optimal subset, it constitutes an effective and low-complexity approach.

The stepwise approach can be employed for antenna selection considering various selection metrics. For example in \cite{gorokhov2003receive,gharavi2004fast,sanayei2004capacity}, iterative stepwise selection algorithms are proposed in which channel capacity was taken as the measure of performance. Simulation results demonstrated that the performance of this algorithm almost captures the optimal performance for moderate number of transmit antennas. The approach is further extended in recent studies, e.g., \cite{zhou2014iterative,bereyhi2018stepwise,gkizeli2014maximum}, considering some other performance metrics such as energy efficiency and receive signal-to-noise ratio. 

Recent studies have demonstrated that antenna selection can be employed as an effective means for secrecy enhancement in massive \ac{mimo} wiretap settings \cite{huang2015secure,asaad2017optimal}. Such studies, however, do not provide algorithmic approaches which exploit this property. In this paper, we develop a stepwise algorithm for antenna selection in massive multiuser \ac{mimo} wiretap settings. Our investigations demonstrate that stepwise antenna selection can considerably enhance the secrecy performance without imposing a computational burden onto the system. 

\hspace*{-1mm}\textit{Notations:} Scalars, vectors and matrices are shown~with~non-bold, bold lower case and bold upper case letters, respectively. The complex plain is shown by $\setC$. $\mH^{\her}$, $\mH^{*}$ and $\mH^{\trp}$ indicate the Hermitian, complex conjugate and transpose of $\mH$, respectively. $\log\left(\cdot\right)$ indicates the binary logarithm, and $\E$ represents the expectation operator. For brevity, we define $[x]^+=\max\{0, x\}$ and abbreviate $\{1, \ldots, N\}$ by $[N]$.
\section{Problem Formulation}
\label{sec:sys}
We consider secure downlink transmission in a massive multiuser \ac{mimo} wiretap setting consisting of a \ac{bs} with $M$ transmit antennas, $K$ single-antenna legitimate receivers and an eavesdropper equipped with $N$ receive antennas. The \ac{bs} is assumed to be equipped with $L_{\max}$ \ac{rf}-chains with $L_{\max} \leq M$. For this setting, uplink channel coefficients from the users to the antenna array at the \ac{bs} are enclosed in the matrix $\mH \in \setC^{M \times K}$. $\mG \in \setC^{M \times N}$ represents the channel from the eavesdropper to the \ac{bs}. The system is assumed to operate in standard \ac{tdd} mode meaning that the channels are reciprocal. The \ac{bs} intends to transmit confidential messages to the users over this wiretap channel while the eavesdropper seeks to recover information conveyed from the \ac{bs} to the legitimate users. The \ac{csi} of the main and the eavesdropper's channel is assumed to be known at the \ac{bs}. 
\subsection{System Model}
At the beginning of each coherence interval, the \ac{bs} selects $L \leq L_{\max}$ transmit antennas based on the \ac{csi} of the main and the eavesdropper's channel. Let $\bs= [s_1, \cdots, s_K]^{\trp}$ be the vector of information symbols. The \ac{bs} precodes $\bs$ linearly as
\begin{align}
\bx=\sqrt{P} \mW_L \bs.
\end{align}
for some $P$. $\mW_L \in \setC^{L \times K}$ is the~signal shaping~matrix which satisfies $\E{\tr{\mW_L \mW_L^\her}}=1$. The subscript $L$ indicates the number of active transmit antennas. Assuming $\E{\bs \bs^{\her}}=\mI_K$, the transmit power  reads $\E{\bx^\her \bx}= P$. We further assume that the transmit power is constrained by $P\leq P_{\max}$.

The precoded signal $\bx\in\setC^L$ is transmitted over the selected antennas. Denoting the indices of the selected antennas with $\setL=\{i_1, \cdots, i_L \}$, the signals received at user terminals read
\begin{align}
\by=\mH_{\setL}^{\trp} \bx + \bn_\mm \label{eq:1}
\end{align}
where $\by=[y_1, \ldots,y_K]^\trp$ with $y_k$ being the received signal at user $k$, $\mH_\setL \in \setC^{L\times K}$ denotes the effective channel enclosing the rows of $\mH$ indexed by $\setL$ and $\bn_\mm$ encloses \ac{iid} zero-mean Gaussian noise at user terminals whose variances are $\sigma_\mm^2$, i.e., $\bn_\mm \sim \mac\man(\boldsymbol{0}, \sigma_\mm^2 \mI_K)$. 

The received signal at the eavesdropper moreover reads
\begin{align}
\bz=\mG_\setL^{\trp} \bx + \bn_\ee  \label{eq:2}
\end{align}
where $\mG_\setL \in \setC^{L\times N}$ is the effective eavesdropper~channel~corresponding to $\setL$ and $\bn_\ee\in\setC^N$ denotes zero-mean complex Gaussian noise with variance $\sigma_\ee^2$, i.e., $\bn_\ee \sim \mac\man( \boldsymbol{0} , \sigma_\ee^2 \mI_{N})$.
%
%
%

\subsection{Secrecy Performance Metric}
From information-theoretic points of view,~the~secrecy~performance is properly quantified via the achievable secrecy rate. For the setting under study, the achievable secrecy rate for user $k$ is given by \cite{oggier2011secrecy, khisti2010secure}
\begin{align}
\mar_k^\rms \left( P,\setL \right) =[\mar_k^\mm \left( P,\setL \right) -\mar_k^\ee \left( P,\setL \right) ]^+.
\end{align}
Here, the arguments $P$ and $\setL$ indicate the dependency on the transmit power and selected antennas. $\mar_k^\mm \left( P,\setL \right)$  denotes the rate to user $k$ achieved over the main channel and $\mar_k^\ee \left( P,\setL \right)$ is the information leakage from user $k$ to the eavesdropper.

Assuming that the \ac{csi}s of the both channels are available at the receiving terminals, the maximum achievable rate for user $k$ over the main channel is lower-bounded by~\cite{caire2010multiuser,alves2012performance}  
\begin{align}
\mar_k^\mm \left( P,\setL \right) =\log \left( 1+\gamma_k^\mm \left( P,\setL \right) \right) \label{eq:Rmm}
\end{align}
where $\gamma_k^\mm \left( P,\setL \right) $ denotes the \ac{sinr} at user $k$ and is given by
\begin{align}
\gamma_k^\mm \left( P,\setL \right) =\dfrac{\rho_\mm t_k(\mH_\setL , \mW_L)}{1+ \rho_\mm u_k(\mH_\setL , \mW_L)}.
\end{align}
Here, $\rho_\mm \coloneqq P/\sigma_\mm^2$ and $t_k(\mH_\setL , \mW_L )$ and $u_k(\mH_\setL , \mW_L)$ are
\begin{subequations}
\begin{align}
t_k(\mH_\setL, \mW_L ) &\coloneqq \abs{\mh_{\setL k}^{\trp}\mw_{Lk}}^2 \\
u_k(\mH_\setL, \mW_L ) &\coloneqq \sum_{j=1, j\ne k}^K \abs{\mh_{\setL k}^{\trp}\mw_{Lj}}^2
\end{align}
\end{subequations}
where $\mh_{\setL j}$ and $\mw_{Lj}$ denote the $j$-th columns of~$\mH_\setL$ and $\mW_L $, respectively. 

The information leakage from user $k$ is upper-bounded by considering the worst-case scenario in which the eavesdropper is able to cancel out all the interfering signals while overhearing the message of user $k$. The maximum information leakage from user $k$ to the eavesdropper is bounded from above as
\begin{align}
\mar_k^\ee \left( P,\setL \right) =\log \left( 1+\gamma_k^\ee \left( P,\setL \right) \right) \label{eq:Ree}
\end{align}
where $\gamma_k^e \left( P,\setL \right)$ is the \ac{sinr} at the eavesdropper while overhearing the message of user $k$ and is given by
\begin{align}
\gamma_k^\ee \left( P,\setL \right) ={\rho_\ee t_k(\mG_\setL, \mW_L )}
\end{align}
with $\rho_\ee = P/\sigma_\ee^2$ and $t_k(\mG_\setL, \mW_L )$ reading
\begin{align}
t_k(\mG_\setL , \mW_L ) &\coloneqq \norm{\mG_\setL^{\trp}\mw_{L k}}^2. 
\end{align}
This bound is tight when other users cooperate with~the~eavesdropper such that it retrieves the interfered signals \cite{zhu2016linear}. 

From \eqref{eq:Rmm} and \eqref{eq:Ree}, one concludes that the~secrecy~rate~achievable for user $k$ is bounded from below by
\begin{align}
\mar_k^\rms \left( P,\setL \right) =\left[ \log \frac{1+\gamma_k^\mm  \left( P,\setL \right) }{1+\gamma_k^\ee  \left( P,\setL \right) } \right]^+ . \label{eq:Rss}
\end{align}
Consequently, the average achievable secrecy rate with respect to the weighting vector $\bw=\left[ w_1, \ldots, w_K \right]$ is given by
\begin{align}
\bar{\mar}^\rms \left( P,\setL \right | \bw)  = \sum_{k=1}^K w_k \mar_k^\rms \left( P,\setL \right).
\end{align}

Throughout the paper, we consider $\bar{\mar}^\rms \left( P,\setL | \bw \right)$ to be~the~secrecy~performance metric of this multiuser setting. Our main objective is to develop an iterative algorithm which effectively selects a subset of transmit antennas and controls the transmit power with respect to this performance metric.

\section{Joint Antenna Selection and Power Control}
We find the optimal power level $P$ and the optimal selection subset $\setL$ for given $\bw$ as 
\begin{align}
\left( P,\setL \right) = \argmax_{\substack{ { 0 \leq Q \leq P_{\max}} \vspace*{2mm}\\ {\setS \subseteq [M], \abs{\setS} \leq L_{\max}}}} \bar{\mar}^\rms \left( Q,\setS | \bw \right). \label{eq:opt}
\end{align}
The combinatorial optimization problem \eqref{eq:opt} is not practical for large $M$. Consequently, one may employ an alternative approach with feasible computational complexity at the expense~of suboptimality. In this section, we develop an iterative algorithm for antenna selection and power control via stepwise regression. For the sake of brevity, we assume that the \ac{bs} employs \ac{mrt} precoding whose signal shaping matrix for $L$ active transmit antennas indexed with $\setL$ is given by
\begin{align}
\mW_L = \beta_{L} \mH_{\setL}^*
\end{align}
with $\beta_{L} \coloneqq  \tr{\mH_\setL \mH_\setL^\her}^{-1/2}$. Nevertheless, the results can be extended to other linear precoding schemes by standard lines of derivations. The extension is briefly discussed later on.

\subsection{\ac{tas} via Stepwise Regression}
In the stepwise approach, the transmit antennas are iteratively selected. Starting from a single active antenna, assume that $\ell < L_{\max}$ antennas have been already selected, and we intend to select the $(\ell+1)$-st transmit antenna. Denoting the index set of $\ell$ selected antennas with $\setL_0$, the set of indices in the next step is $\setL_1 = \setL_0 \cup \{ i_{\ell+1} \}$ where $i_{\ell+1}$ denotes the index of the transmit antenna being selected in step $\ell+1$. The effective channels and the signal shaping matrix read
\begin{subequations}
\begin{align}
\mH_{\setL_1}^\trp &= \left[ \mH_{\setL_0}^\trp, \ \mh_{\ell+1} \right] \label{eq:channel} \\
\mG_{\setL_1}^\trp &= \left[ \mG_{\setL_0}^\trp, \ \bg_{\ell+1} \right] \label{eq:channel} \\
\mW_{\ell +1}^\trp &= \alpha \left( i_{\ell +1} \right) \left[ \mW_{\ell}^\trp, \ \beta_{\ell} \mh_{\ell+1}^* \right] \label{eq:precoder}
\end{align}
\end{subequations}
where $\mh_{\ell+1}=[h_1,\ldots,h_K]^\trp$ and $\bg_{\ell+1}$ are the column vectors in $\mH^\trp$ and $\mG^\trp$ indexed by $i_{\ell+1}$, and $\alpha \left( i_{\ell +1} \right) \coloneqq \beta_{\ell +1} / \beta_{\ell}$ is determined as
\begin{align}
\alpha \left( i_{\ell +1} \right) = {1}/{\sqrt{1 + \beta_{\ell}^2 \norm{\bh_{\ell+1}}^2 }}.
\end{align}
Moreover, $\mH_{\setL_0}$ and $\mG_{\setL_0}$ denote the effective uplink channels in step $\ell$, $\mH_{\setL_1}$ and $\mG_{\setL_1}$ are the effective channels in step $\ell+1$, and $\mW_{\ell}$ and $\mW_{\ell+1}$ represent the \ac{mrt} signal shaping matrices before and after selecting the new antenna, respectively. 

Considering \eqref{eq:channel}-\eqref{eq:precoder}, the performance of the setting in step $\ell+1$ is described as a stepwise update of the performance in step $\ell$. To illustrate this statement, assume fixed power $P$ at the transmitter. In this case, one can write $\gamma_k^\mm \left( P,\setL_1 \right)$ and $\gamma_k^\ee \left( P,\setL_1 \right)$ in terms of the \ac{sinr} in step $\ell$ as
\begin{subequations}
\begin{align}
1+\gamma^\mm \left( P,\setL_1 \right) &= \theta_k^\mm \left( P, i_{\ell+1}  \right) \left( 1+ \gamma^\mm \left( P,\setL_0 \right) \right) \\
1+\gamma^\ee \left( P,\setL_1 \right) &= \theta_k^\ee \left( P, i_{\ell+1}  \right) \left( 1+ \gamma^\ee \left( P,\setL_0 \right) \right)
\end{align}
\end{subequations}
where $\theta_k^{\mm} \left( P, \ell+1  \right)$ and $\theta_k^{\ee} \left( P, \ell+1  \right)$ are given by 
\begin{subequations}
\begin{align}
\theta_k^{\mm} \left( P, i_{\ell+1}  \right) &= \frac{\alpha^2 \left( i_{\ell+1}  \right) + \epsilon^{\mm}_k \left( P, i_{\ell+1}  \right)}{\alpha^2 \left( i_{\ell+1}  \right) + \psi^{\mm}_k \left( P, i_{\ell+1}  \right)} \\
\theta_k^\ee \left( P, i_{\ell+1}  \right) &= \alpha^2 \left( i_{\ell+1}  \right) + \epsilon^\ee_k \left( P, i_{\ell+1}  \right)
\end{align}
\end{subequations}
where $\epsilon^{\mm}_k \left( P, i_{\ell+1}  \right)$, $\psi^{\mm}_k \left( P, i_{\ell+1}  \right)$ and $\epsilon^{\ee}_k \left( P, i_{\ell+1}  \right)$ are given by \eqref{eq:endol}-\eqref{eq:endel} on the top of the next page.
\begin{figure*}[!t]
\begin{subequations}
\begin{align}
\epsilon^\mm_k \left( P, i_{\ell+1}  \right) &= \frac{1+\rho_\mm \alpha^2 \left( i_{\ell+1}  \right) \beta_\ell \left( \sum_{j=1}^K \beta_\ell \abs{h_k h^*_j}^2 + 2 \re{ \mh^\trp_{\setL_0 k} \mw_{\ell j} h_k h_j^* } \right) -\alpha^2 \left( i_{\ell+1}  \right)  }{1+ \rho_\mm \left( t_k(\mH_{\setL_0} , \mW_\ell ) +u_k(\mH_{\setL_0} , \mW_\ell ) \right) } \label{eq:endol} \\
\psi^\mm_k \left( P, i_{\ell+1}  \right) &= \frac{1+\rho_\mm \alpha^2 \left( i_{\ell+1}  \right) \beta_\ell \left( \sum_{j=1, j\ne k}^K \beta_\ell \abs{h_k h^*_j}^2 + 2 \re{ \mh^\trp_{\setL_0 k} \mw_{\ell j} h_k h_j^* }\right) -\alpha^2 \left( i_{\ell+1}  \right)  }{1+ \rho_\mm u_k(\mH_{\setL_0} , \mW_\ell ) } \\
\epsilon^\ee_k \left( P, i_{\ell+1}  \right) &= \frac{1+\rho_\ee \alpha^2 \left( i_{\ell+1}  \right) \beta_\ell \left( \beta_\ell \abs{h_k}^2 \norm{\bg_{\ell+1}}^2 + 2 \re{ h_k \bg_{\ell+1}^\her \mG^\trp_{\setL_0} \mw_{\ell k} } \right) -\alpha^2 \left( i_{\ell+1}  \right)  }{1+ \rho_\ee t_k(\mG_{\setL_0} , \mW_\ell )  } \label{eq:endel}
\end{align}
\end{subequations} 
\centering\rule{17cm}{0.1pt}
\vspace*{2pt}
\end{figure*}

Consequently, the average achievable secrecy rate in step $\ell+1$, i.e., $\bar{\mar}^\rms \left( P,\setL_1 | \bw \right)$, can be written as
\begin{align}
\bar{\mar}^\rms \left( P,\setL_1 | \bw\right) = \bar{\mar}^\rms \left( P,\setL_0 | \bw \right) + \Theta \left( P,i_{\ell+1} | \bw \right) \label{eq:update}
\end{align}
where $\Theta \left( P,i_{\ell+1} | \bw \right)$ is defined as
\begin{align}
\Theta \left( P,i_{\ell+1} | \bw \right) \coloneqq \sum_{k=1}^K w_k \log \frac{\theta_k^\mm \left( P, i_{\ell+1}  \right)}{\theta_k^\ee \left( P, i_{\ell+1}  \right)}.
\end{align}

From \eqref{eq:update}, one observes that the performance metric in step $\ell +1 $ is given by an update of the metric in step~$\ell$~via a single term depending on $i_{\ell+1}$. Stepwise regression suggests that in each step, we select the transmit antenna which maximizes this single update term. In this case, the active antennas are selected such that the growth in the performance is optimized in \textit{each step}. In contrast to optimal \ac{tas}, this stepwise approach has linear complexity which is computationally feasible in practice. Nevertheless, one should note that it does not necessarily lead to the globally optimal solution given by \eqref{eq:opt}. 
\subsection{Iterative \ac{tas} and Power Control Algorithm}
We develop an iterative algorithm for joint power control and \ac{tas} in this section. The algorithm employs the stepwise \ac{tas} approach while iteratively updating the transmit power in each step. It is given in Algorithm~\ref{alg1} and its details~are~illustrated in the sequel.
\subsubsection*{Initialization}
For a given $\bw$, the algorithm starts with the following initialization: 
\begin{itemize}
\item The index of the first active antenna is set to $i_1$ such that
\begin{align}
i_1 = \argmax_{i \in[M]} \frac{\norm{\mH_{\{i\}}}}{\norm{\mG_{\{i\}}}}.
\end{align}
\item The transmit power is set to $P_1$ such that the average achievable secrecy rate for $\setL=\{i_1\}$ is maximized.
\end{itemize}

\subsubsection*{Iterative \ac{tas}}
At step $\ell\in[L_{\max}]$, the algorithm selects the transmit antenna indexed with $i_{\ell+1}$ from the non-selected antennas such that the growth term $\Theta \left( P_\ell , i_{\ell+1} | \bw \right)$ is maximized where $P_\ell$ is the transmit power being set at the end of step $\ell$.

\subsubsection*{Iterative Power Control}
The transmit power is updated in each iteration after antenna selection such that the average secrecy rate, achieved via the selected antennas, is optimized with respect to $P$. This means that in step $\ell$, after selection of the $(\ell+1)$-st transmit antenna, the selection subset is expanded by $\setL=\setL \cup \{i_{\ell+1} \}$ and the power is updated as
\begin{align}
P_{\ell+1}=\argmax_{0\leq P \leq P_{\max}} \bar{\mar}^\rms \left( P,\setL | \bw\right).
\end{align}
\subsubsection*{Stopping Criteria}
When the performance metric monotonically increases with respect to the number of selected antennas, the stepwise selection is continued until $L_{\max}$ transmit antennas are set active. There exist, however, scenarios for which the increase in number of active antennas does not necessarily enhance the performance metric \cite{asaad2017optimal}. In this case, the optimal stepwise update term, i.e., $\Theta \left( P_\ell , i_{\ell+1} | \bw \right)$, does not return a positive value after some iterations. We therefore stop the algorithm either when the number of active antennas is $L_{\max}$ or when the optimal stepwise update term is non-positive, i.e., $\Theta( P_\ell , i_{\ell +1} |\bw ) \leq 0$ the latter criteria is labeled by $\sf STC$ in Algorithm~\ref{alg1}.

\begin{algorithm}[t]
\caption{Iterative Joint \ac{tas} and Power Control}
\label{alg1}
\begin{algorithmic}[0]
\In Channel matrices $\mH$ and $\mG$, and $P_{\max}$, $L_{\max}$ and $\bw$
\Initiate Let $\ell=1$ and
\begin{align}
i_1 = \argmax_{i \in[M]} \frac{\norm{\mH_{ \{i \} } }}{\norm{\mG_{ \{i \} }}}.
\end{align}
Set $\setL=\{ i_1 \}$, $\mH_{\setL} = \mH(i_1, :)$, $\mG_{\setL} = \mG(i_1, :)$ and
\begin{align}
P_1=\argmax_{0\leq P \leq P_{\max}} \bar{\mar}^\rms \left( P,\setL |\bw\right).
\end{align}
\While\NoDo $\ell< L_{\max}$
\begin{align}
i_{\ell +1} &= \argmax_{i\in[M]\backslash\setL } \Theta(P_\ell , i |\bw ).
\end{align}
\If{$\Theta( P_\ell , i_{\ell +1} |\bw ) \leq 0$} \hfill $\sf STC$

break
\EndIf \vspace*{1.5mm}\\

Set $\mH_{\setL}= \left[\mH_\setL^\trp, \hspace*{1mm} \mH(i_{\ell+1}, :)^\trp \right]^\trp$ 
and update the precoder as 
\begin{align}
\mW_{\ell+1}= \alpha_{\ell+1} \begin{bmatrix}
          \mW_{\ell} \\ 
         \beta_\ell \hspace*{.4mm}\mH^*(i_{\ell+1}, :)
         \end{bmatrix}.
\end{align}
Update $\setL = \setL \cup \set{i_{\ell+1}}$ %
and the transmit power as
\begin{align}
P_{\ell+1}=\argmax_{0\leq P \leq P_{\max}} \bar{\mar}^\rms \left( P,\setL |\bw \right).
\end{align}
Set $\ell=\ell+1$.\vspace*{1mm}
\EndWhile \vspace*{2mm}
\Out $L =\ell$, $P=P_\ell$ and $\setL$.
\end{algorithmic}
\end{algorithm}

\subsection{Further Extensions}
Although the results have been derived for \ac{mrt} precoding and average secrecy rate, the approach can be extended~to other linear precoders and performance metrics; see discussions~in \cite{bereyhi2018stepwise}. For other linear precoders, the rank-one updates, similar to the one derived for \ac{mrt} precoding in \eqref{eq:precoder}, are derived using the Sherman-Morrison formula \cite{bartlett1951inverse}. By similar lines of derivations, the stepwise update rule is extended to multiple performance metrics. Due to lack of space, further derivations are skipped and left for the extended version of the manuscript.

\section{Numerical Investigations}
\begin{figure}[t]
\hspace*{-1.3cm}  
\resizebox{1.3\linewidth}{!}{

\pstool[width=.35\linewidth]{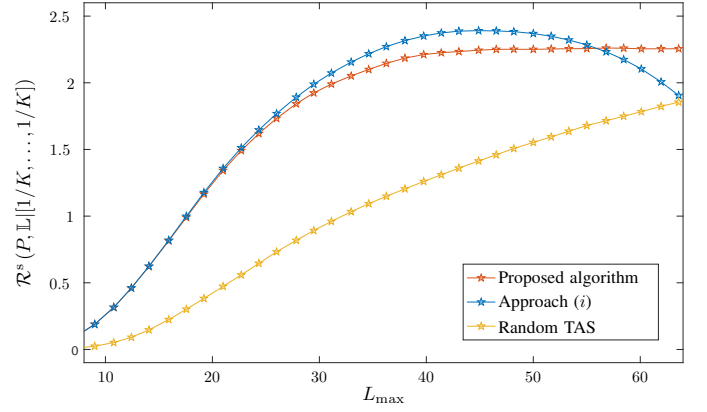}{

\psfrag{L}[c][c][0.22]{$L_{\max}$}
\psfrag{rate}[c][t][0.22]{$\mar^{\rm s} \left( P, \setL | [1/K, \ldots,1/K ] \right)$}
\psfrag{ALGMITCONSAAA}[l][l][0.2]{Proposed algorithm}
\psfrag{ALGOHICONSAAA}[l][l][0.2]{Approach ($i$)}
\psfrag{ALGMITRNDSAAA}[l][l][0.2]{Random \ac{tas}}

\psfrag{1}[r][c][0.18]{$1$}
\psfrag{2}[r][c][0.18]{$2$}
\psfrag{0.5}[r][c][0.18]{$0.5$}
\psfrag{1.5}[r][c][0.18]{$1.5$}
\psfrag{2.5}[r][c][0.18]{$2.5$}

\psfrag{40}[c][c][0.18]{$40$}
\psfrag{10}[c][c][0.18]{$10$}
\psfrag{50}[c][c][0.18]{$50$}
\psfrag{20}[c][c][0.18]{$20$}
\psfrag{60}[c][c][0.18]{$60$}
\psfrag{30}[c][c][0.18]{$30$}
}
}
\vspace*{-8mm}\caption{Performance of the \ac{tas} approaches for $M=64$, $K=4$, $P_{\max}=1$, $\sigma_\mm^2=\sigma_\ee^2=0.1$ and $N=8$. The proposed algorithm stops selecting antennas at $L=37$, since further selection degrades the performance. Such degradation is observed in approach~($i$).\vspace*{-3mm}}
\label{fig}
\end{figure}
We investigate the proposed algorithm numerically by considering the following sample setting: The \ac{bs} has a transmit antenna array of size $M=64$ and $L_{\max}$ \ac{rf}-chains. Moreover, the eavesdropper is equipped with $N=8$ receive antennas. The number of users is set to $K=4$. For simplicity, the main channel and the eavesdropper's channel are assumed to be \ac{iid} unit-variance Rayleigh fading meaning that their entries are \ac{iid} zero-mean and unit-variance complex Gaussian random variables. The noise variances at the user terminals and the eavesdropper are set to $\sigma_\mm^2=\sigma_\ee^2=0.1$, and the transmit power $P$ is constrained by $P_{\max}=1$. The weighting factors $w_k$ are set to $1/K$ for all the users.

In Fig.~\ref{fig}, the average achievable secrecy rate $\bar{\mar}^\rms \left( P,\setL |\bw \right)$ is given as a function of the number of \ac{rf}-chains $L_{\max}$ for three different approaches: 
\begin{inparaenum}
\item[($i$)] The stepwise approach given in Algorithm~\ref{alg1} without the stopping criteria $\sf STC$.
\item[($ii$)] The proposed iterative algorithm with the stopping criteria $\sf STC$.
\item[($iii$)] Random \ac{tas}.
\end{inparaenum}
As the figure depicts, the secrecy rate is not an increasing function of $L_{\max}$ in the stepwise approach. Such an observation is also reported in \cite{asaad2017optimal} via large-system analyses. The optimal choice for the number of active transmit antennas is some $L < L_{\max}$ which is approximated by the proposed iterative approach. As the figure depicts, the proposed algorithm stops selecting antennas around $L=37$, due to the fact that further selection degrades the performance. The slight degradation in the performance of the algorithm with the stopping criteria $\sf STC$ is due to fact that the algorithm solves the coupled problems of power control and \ac{tas} separately. For the sake of comparison, we have also evaluated the performance of random \ac{tas} for this setting. The figure shows a significantly degraded performance.

\section{Conclusion and Outlook}
The proposed iterative algorithm for joint \ac{tas} and power control in massive \ac{mimo} wiretap settings selects the active transmit antennas using the forward selection method from stepwise regression. The proposed algorithm significantly enhances the secrecy performance while enjoying low computational complexity.

The large-system performance characterization of the proposed algorithm is an interesting direction for future work. The work in this direction is currently ongoing.

\bibliography{ref}

\begin{thebibliography}{10}
\providecommand{\url}[1]{#1}
\csname url@samestyle\endcsname
\providecommand{\newblock}{\relax}
\providecommand{\bibinfo}[2]{#2}
\providecommand{\BIBentrySTDinterwordspacing}{\spaceskip=0pt\relax}
\providecommand{\BIBentryALTinterwordstretchfactor}{4}
\providecommand{\BIBentryALTinterwordspacing}{\spaceskip=\fontdimen2\font plus
\BIBentryALTinterwordstretchfactor\fontdimen3\font minus
  \fontdimen4\font\relax}
\providecommand{\BIBforeignlanguage}[2]{{%
\expandafter\ifx\csname l@#1\endcsname\relax
\typeout{** WARNING: IEEEtran.bst: No hyphenation pattern has been}%
\typeout{** loaded for the language `#1'. Using the pattern for}%
\typeout{** the default language instead.}%
\else
\language=\csname l@#1\endcsname
\fi
#2}}
\providecommand{\BIBdecl}{\relax}
\BIBdecl

\bibitem{khisti2010secure}
A.~Khisti and G.~W. Wornell, ``Secure transmission with multiple
  antennas—{P}art {II}: The {MIMOME} wiretap channel,'' \emph{IEEE Transactions on
  Information Theory}, vol.~56, no.~11, pp. 5515--5532, 2010.

\bibitem{oggier2011secrecy}
F.~Oggier and B.~Hassibi, ``The secrecy capacity of the {MIMO} wiretap
  channel,'' \emph{IEEE Trans. on Inf. Theory}, vol.~57, no.~8, pp. 4961--4972,
  2011.

\bibitem{liu2009note}
T.~Liu and S.~Shamai, ``A note on the secrecy capacity of the multiple-antenna
  wiretap channel,'' \emph{IEEE Transactions on Information Theory}, vol.~55,
  no.~6, pp. 2547--2553, 2009.

\bibitem{kapetanovic2015physical}
D.~Kapetanovic, G.~Zheng, and F.~Rusek, ``Physical layer security for massive
  MIMO: An overview on passive eavesdropping and active attacks,'' \emph{IEEE
  Communications Magazine}, vol.~53, no.~6, pp. 21--27, 2015.

\bibitem{molisch2005capacity}
A.~F. Molisch, M.~Z. Win, Y.-S. Choi, and J.~H. Winters, ``Capacity of {MIMO}
  systems with antenna selection,'' \emph{IEEE Transactions on Wireless
  Communications}, vol.~4, no.~4, pp. 1759--1772, 2005.

\bibitem{asaad2017tas}
S.~Asaad, A.~Bereyhi, R.~R. M{\"u}ller, and A.~M. Rabiei, ``Asymptotics of
  transmit antenna selection: Impact of multiple receive antennas,''
  \emph{IEEE International Conference on Communications (ICC)}, 2017.

\bibitem{sedaghat2016load}
M.~A. Sedaghat, V.~I. Barousis, R.~R. M{\"u}ller, and C.~B. Papadias, ``Load
  modulated arrays: a low-complexity antenna,'' \emph{IEEE Communications
  Magazine}, vol.~54, no.~3, pp. 46--52, 2016.

\bibitem{liang2014low}
L.~Liang, W.~Xu, and X.~Dong, ``Low-complexity hybrid precoding in massive
  multiuser MIMO systems,'' \emph{IEEE Wireless Communications Letters},
  vol.~3, no.~6, pp. 653--656, 2014.

\bibitem{sanayei2004antenna}
S.~Sanayei and A.~Nosratinia, ``Antenna selection in {MIMO} systems,''
  \emph{IEEE Communications Magazine}, vol.~42, no.~10, pp. 68--73, 2004.

\bibitem{bereyhi2017asymptotics}
A.~Bereyhi, M.~A. Sedaghat, and R.~R. M{\"u}ller, ``Asymptotics of nonlinear
  LSE precoders with applications to transmit antenna selection,'' in
  \emph{IEEE International Symposium on Information Theory (ISIT)}, pp.~81--85, 2017.

\bibitem{bereyhi2017nonlinear}
A.~Bereyhi, M.~A. Sedaghat, S.~Asaad, and R.~R. M{\"u}ller, ``Nonlinear
  precoders for massive MIMO systems with general constraints,''
  \emph{International ITG Workshop on Smart Antennas (WSA)}, 2017.

\bibitem{marzetta2010noncooperative}
T.~L. Marzetta, ``Noncooperative cellular wireless with unlimited numbers of
  base station antennas,'' \emph{IEEE Transactions on Wireless Communications},
  vol.~9, no.~11, pp. 3590--3600, 2010.

\bibitem{bereyhi2018stepwise}
A.~Bereyhi, S.~Asaad, and R.~R. M{\"u}ller, ``Stepwise transmit antenna
  selection in downlink massive multiuser MIMO,'' \emph{International ITG
  Workshop on Smart Antennas (WSA); available on arXiv, arXiv:1802.05148}, 2018.

\bibitem{duda2012pattern}
R.~O. Duda, P.~E. Hart, and D.~G. Stork, \emph{Pattern classification}.\hskip
  1em plus 0.5em minus 0.4em\relax John Wiley \& Sons, 2012.

\bibitem{han2011data}
J.~Han, J.~Pei, and M.~Kamber, \emph{Data mining: concepts and
  techniques}. Elsevier, 2011.

\bibitem{gorokhov2003receive}
A.~Gorokhov, D.~A. Gore, and A.~J. Paulraj, ``Receive antenna selection for
  {MIMO} spatial multiplexing: theory and algorithms,'' \emph{IEEE Transactions
  on Signal Processing}, vol.~51, no.~11, pp. 2796--2807, 2003.

\bibitem{gharavi2004fast}
M.~Gharavi-Alkhansari and A.~B. Gershman, ``Fast antenna subset selection in
  {MIMO} systems,'' \emph{IEEE Transactions on Signal Processing}, vol.~52,
  no.~2, pp. 339--347, 2004.

\bibitem{sanayei2004capacity}
S.~Sanayei and A.~Nosratinia, ``Capacity maximizing algorithms for joint
  transmit-receive antenna selection,'' in \emph{$38^{\rm th}$ Asilomar Conference on Signals, Systems and Computers}, vol.~2, pp.~1773--1776, 2004

\bibitem{zhou2014iterative}
X.~Zhou, B.~Bai, and W.~Chen, ``An iterative algorithm for joint antenna
  selection and power adaptation in energy efficient MIMO,'' in
  \emph{IEEE International Conference on Communications (ICC)}, pp. 3812--3816, 2014.

\bibitem{gkizeli2014maximum}
M.~Gkizeli and G.~N. Karystinos, ``Maximum-SNR antenna selection among a large
  number of transmit antennas,'' \emph{IEEE Journal of Selected Topics in
  Signal Processing}, vol.~8, no.~5, pp. 891--901, 2014.

\bibitem{huang2015secure}
Y.~Huang, F.~S. Al-Qahtani, T.~Q. Duong, and J.~Wang, ``Secure transmission in
  MIMO wiretap channels using general-order transmit antenna selection with
  outdated csi,'' \emph{IEEE Transactions on Communications}, vol.~63, no.~8,
  pp. 2959--2971, 2015.

\bibitem{asaad2017optimal}
S.~Asaad, A.~Bereyhi, R.~R. M{\"u}ller, R.~F. Schaefer, and A.~M. Rabiei,
  ``Optimal number of transmit antennas for secrecy enhancement in massive
  {MIMOME} channels,'' \emph{IEEE Global Communications Conference (GLOBECOM)}, 2017.

\bibitem{caire2010multiuser}
G.~Caire, N.~Jindal, M.~Kobayashi, and N.~Ravindran, ``Multiuser {MIMO}
  achievable rates with downlink training and channel state feedback,''
  \emph{IEEE Transactions on Information Theory}, vol.~56, no.~6, pp. 2845--2866, 2010.

\bibitem{alves2012performance}
H.~Alves, R.~D. Souza, M.~Debbah, and M.~Bennis, ``Performance of transmit
  antenna selection physical layer security schemes,'' \emph{IEEE Signal
  Processing Letters}, vol.~19, no.~6, pp. 372--375, 2012.

\bibitem{zhu2016linear}
J.~Zhu, R.~Schober, and V.~K. Bhargava, ``Linear precoding of data and
  artificial noise in secure massive {MIMO} systems,'' \emph{IEEE Transactions
  on Wireless Communications}, vol.~15, no.~3, pp. 2245--2261, 2016.

\bibitem{bartlett1951inverse}
M.~S. Bartlett, ``An inverse matrix adjustment arising in discriminant
  analysis,'' \emph{The Annals of Mathematical Statistics}, vol.~22, no.~1, pp.
  107--111, 1951.

\end{thebibliography}
\bibliographystyle{IEEEtran}

\begin{acronym}
\acro{tdd}[TDD]{Time Division Duplexing}
\acro{mimo}[MIMO]{Multiple-Input Multiple-Output}
\acro{misome}[MISOME]{Multiple-Input Single-Output Multiple-Eavesdropper}
\acro{csi}[CSI]{Channel State Information}
\acro{awgn}[AWGN]{Additive White Gaussian Noise}
\acro{iid}[i.i.d.]{independent and identically distributed}
\acro{ut}[UT]{User Terminal}
\acro{bs}[BS]{Base Station}
\acro{mt}[MT]{Mobile Terminal}
\acro{eve}[Eve]{Eavesdropper}
\acro{tas}[TAS]{Transmit Antenna Selection}
\acro{lse}[LSE]{Least Squared Error}
\acro{rhs}[r.h.s.]{right hand side}
\acro{lhs}[l.h.s.]{left hand side}
\acro{wrt}[w.r.t.]{with respect to}
\acro{rs}[RS]{Replica Symmetry}
\acro{rsb}[RSB]{Replica Symmetry Breaking}
\acro{papr}[PAPR]{Peak-to-Average Power Ratio}
\acro{mrt}[MRT]{Maximum Ratio Transmission}
\acro{zf}[ZF]{Zero Forcing}
\acro{rzf}[RZF]{Regularized Zero Forcing}
\acro{snr}[SNR]{Signal to Noise Ratio}
\acro{sinr}[SINR]{Signal-to-Interference-plus-Noise Ratio}
\acro{rf}[RF]{Radio Frequency}
\acro{mf}[MF]{Match Filtering}
\acro{mmse}[MMSE]{Minimum Mean Squared Error}
\end{acronym}

\end{document}